\documentclass[a4,11pt,pcxf,epsf,amssymb]{article}
\usepackage{amssymb}
\usepackage{color}
\usepackage{graphicx}
\usepackage{chicago}
\usepackage{hyperref}
\usepackage{amsmath,amsfonts,latexsym,amssymb}
\usepackage{bm}
\begin{document}
\title{\bf Extending approximate Bayesian computation methods to high dimensions via a Gaussian copula model}
\author{J. Li\footnote{Department of Statistics and Applied Probability,
National University of Singapore, Singapore 117546.}, \:
D. J. Nott$^*$, Y. Fan\footnote{School of Mathematics and Statistics, University of New South 
Wales, Sydney 2052 Australia.}\:\: and S. A. Sisson$^\dagger$\footnote{Corresponding Author: Email \tt{Scott.Sisson@unsw.edu.au}}}
\date{}

\maketitle

\baselineskip1.9em

\begin{abstract}
\noindent  
Approximate Bayesian computation (ABC) refers to a family of inference methods used in the Bayesian analysis of complex models where evaluation of the likelihood is difficult.  Conventional ABC methods often suffer from the curse of dimensionality, and a marginal adjustment strategy was recently introduced in the literature to improve the performance of ABC algorithms in high-dimensional problems. The marginal adjustment approach is extended using a Gaussian copula approximation. The method first estimates the bivariate posterior for each pair of parameters separately using a $2$-dimensional Gaussian copula, and then combines these estimates together to estimate the joint posterior. The approximation works well in large sample settings when the posterior is approximately normal, but also works well in many cases which are far from that situation due to the nonparametric estimation of the marginal posterior distributions. If each bivariate posterior distribution can be well estimated with a low-dimensional ABC analysis then this Gaussian copula method can extend ABC methods to problems of high dimension. The method also results in an analytic expression for the approximate posterior which is useful for many purposes such as approximation of the likelihood itself.  This method is illustrated with several examples.
\vspace{2mm}

\noindent{\bf Keywords}: Approximate Bayesian Computation (ABC), Gaussian copula, Likelihood free inference, Marginal adjustment, Regression adjustment ABC.
\end{abstract}


\section{Introduction}

Part of the class of ``likelihood-free'' techniques, approximate Bayesian computation (ABC) methods are commonly implemented to draw samples from an approximation to the posterior distribution when the likelihood function is computationally intractable. This scenario arises in an increasingly broad range of discipline areas \shortcite{beaumont+zb02,bortot+cs07,drovandi+p11}.

Denote the prior for a parameter vector $\theta = (\theta_1,\cdots,\theta_p)^\top\in\Theta^p$ as $p(\theta)$, the computationally intractable likelihood function as $L(y|\theta)$, and the resulting posterior distribution as $\pi(\theta|y_{obs})\propto L(y_{obs}|\theta)p(\theta)$, for observed data $y_{obs}$.  The same basic mechanism underlies most ABC algorithms. For each of $i=1,\ldots,N$ candidate draws from the prior distribution, $\theta^{(i)}\sim p(\theta)$, an auxiliary dataset $y^{(i)}\sim L(y|\theta^{(i)})$ is sampled from the data generation process given $\theta^{(i)}$. Suppose that $s=S(y)$ is a vector of summary statistics with $\dim(s)\leq\dim(y)$, and that $s_{obs}=S(y_{obs})$. If $\|s^{(i)}-s_{obs}\|$ is small, for some distance measure $\|\cdot\|$, then $\theta^{(i)}$ could credibly have generated the observed summary data $s_{obs}$, so $\theta^{(i)}$ is 
a possible draw from $\pi(\theta|y_{obs})$. Conversely, if $\|s^{(i)}-s_{obs}\|$ is large,  then $\theta^{(i)}$ is unlikely to have  generated the observed data, so $\theta^{(i)}$ is 
not likely to be a draw from the posterior. Specifically, the resulting samples $(\theta^{(i)}, s^{(i)})$ are draws from the joint distribution
\begin{equation}
\label{eqn:ABCjointPosterior}
	\pi_h^{ABC}(\theta,s|s_{obs})\propto K_h(\|s-s_{obs}\|)L(s|\theta)p(\theta),
\end{equation}
where $K_h$ is a standard smoothing kernel with scale parameter $h>0$. A simple importance sampling ABC algorithm describing this simulation process is given in Table \ref{eqn:importanceSampling}. Note that direct evaluation of the intractable likelihood function is circumvented.

\begin{table}[h]
\centering
\begin{tabular}{cl}
\hline
\\
&{\bf Input:} \\&An observed dataset, $y_{obs}$. \\&A desired number of samples $N>0$.\\ &An importance sampling distribution $f(\theta)$, with $f(\theta)>0$ if $p(\theta)>0$.\\
 &A smoothing kernel $K_h$ and scale parameter $h>0$. \\&A low-dimensional vector of summary statistics $s=S(y)$.\\ &Compute $s_{obs}=S(y_{obs})$.\\
\\
&{\bf Iterate:} \\
& For $i=1,\ldots,N$:\\
1. & Sample a parameter vector from importance distribution $\theta^{(i)}\sim f(\theta)$.\\
2. & Simulate a dataset from the likelihood $y^{(i)} \sim L(y | \theta^{(i)})$ given parameter vector $\theta^{(i)}$.\\
3. & Compute the summary statistics $s^{(i)}=S(y^{(i)})$.\\
4. & Weight each sample $\theta^{(i)}$ by $w^{(i)}\propto K_h(\|s^{(i)}-s_{\text{obs}}\|)p(\theta^{(i)})/f(\theta^{(i)})$.\\
\\
&{\bf Output:}\\
& A set of $i=1,\ldots,N$ samples $(\theta^{(i)},s^{(i)})$ with weights $w^{(i)}$, drawn from $\pi_h^{ABC}(\theta,s|s_{obs})$.\\
\\
\hline
\end{tabular}
\caption{A simple ABC importance sampling algorithm.}\label{eqn:importanceSampling}
\end{table}

Integrating out the auxiliary summary dataset from (\ref{eqn:ABCjointPosterior}) results in the ABC approximation to the posterior
\begin{equation}
\label{eqn:ABCposterior}
	\pi_h^{ABC}(\theta|s_{obs})\propto \int K_h(\|s-s_{obs}\|)L(s|\theta)p(\theta) ds.
\end{equation}
This distribution has the property that if $S(y)$ is sufficient for $\theta$, and if $h\rightarrow 0$ then $\lim_{h\rightarrow 0}\pi_h^{ABC}(\theta|s_{obs})=\pi(\theta|y_{obs})$, so that the exact posterior distribution is recovered. However, in practice sufficient statistics are typically unavailable for intractable models, and in simulations $s^{(i)}\neq s_{obs}$ (so that $h>0$) in all but trivial settings. As a result, $\pi_h^{ABC}(\theta|s_{obs})$ will only approximate the posterior in general. For further details on ABC models and alternative sampling algorithms  see e.g. \shortciteN{beaumont+cmr09,sisson+ft07,marjoram+mpt03,drovandi2011estimation}.

One of the primary restrictions in the application of ABC methods in general is that they suffer from the curse of dimensionality \cite{blum10}. Casual inspection of (\ref{eqn:ABCposterior}) indicates that ABC methods are based on a kernel density estimate of the likelihood function. Kernel density estimation is well known to be reliable only in low dimensions. Here the relevant dimension is in the comparison of $s$ with $s_{obs}$ (not to be confused with the univariate quantity $\|s-s_{obs}\|$). As $\dim(s)\geq\dim(\theta)$ is required for reasons of parameter identifiability, this means that ABC methods perform poorly in models with even a moderate number of parameters. In practice, it is not uncommon that $\dim(s)>>\dim(\theta)$ \shortcite{allinghamkm09,bortot+cs07}, so $\pi_h^{ABC}(\theta|s_{obs})$ can be a poor approximation of $\pi(\theta|y_{obs})$ even for low dimensional models.

In some circumstances, the curse of dimensionality problem can be circumvented. This can occur where the intractable likelihood function factorises in some way  (\shortciteNP{bazin+db10,white+kp15,barthelme+c14}).   For example, suppose that $L(y|\theta)=\prod_j L(y_{(j)}|\theta)$ where $y_{(j)}$ represents some subset of $y$, and that conditional simulation from $L(y_{(j)}|\theta)$ is possible;  in this case, the comparison of $s$ and $s_{obs}$ can  be directly reduced to multiple  lower dimensional (even univariate) comparisons. However, these approaches are problem specific, and are not suitable for usage with general, non-factorisable models.

More generally applicable methods have been proposed, such as the regression and marginal adjustments  \shortcite{beaumont+zb02,nott+fms14}. The regression adjustment takes advantage of the lack of an exact match between $s^{(i)}$ and $s_{obs}$ by constructing a regression model 
 to capture the relationship between the parameter vector and the summary statistics.  \shortciteN{beaumont+zb02} introduced the weighted linear regression model
\begin{eqnarray*} 
	\theta^{(i)}=\alpha+{\beta}^{\top}(s^{(i)}-s_{obs})+\varepsilon_i,
\end{eqnarray*}
where $\alpha$ is a $p\times 1$ vector, $\beta$ is a $q\times p$ matrix of regression coefficients (where $q=\dim(s)$) and $\varepsilon_i$ are zero-mean iid errors, and where
the weight for the pair $(\theta^{(i)},s^{(i)})$ is given by $K_h(\|s^{(i)}-s_{obs}\|)$. 
Writing the least squares estimates of $\alpha$ and $\beta$ as $\hat{\alpha}$ and $\hat{\beta}$, and the resulting empirical residuals as $\hat{\varepsilon}_i$, the linear regression adjusted vector
\begin{eqnarray*}
	\theta^{(i)*}=\theta^{(i)} -{\hat{\beta}}^{\top}(s^{(i)}-s_{obs})=\hat{\alpha}+\hat{\varepsilon}_{i}
\end{eqnarray*}
is approximately a draw from $\pi(\theta|s_{obs})=\lim_{h\rightarrow 0}\pi_h^{ABC}(\theta|s_{obs})$  if the assumptions of the regression model hold.  More flexible non-linear, heteroscedastic regression adjustments have been developed \shortcite{blum+f10,blum+nps13}. The regression adjustment can work well in improving the ABC posterior approximation, however it only mitigates, rather than removes the underlying curse of dimensionality problem \shortcite{nott+fms14}.

The marginal adjustment \shortcite{nott+fms14} 
first constructs estimates of $\pi(\theta|s_{obs})$ using regular ABC with regression adjustment, and precise estimates of the univariate marginal posterior distributions $\pi(\theta_i|s_{obs,(i)})$ for $i=1,\ldots,p$, where $s_{obs,(i)}$ is a subset of $s_{obs}$ informative for $\theta_i$.
The marginal posterior of $\theta_i$ can often be estimated well,  due to the reduced dimensionality of the marginal summary statistic $s_{obs,(i)}$. 
The marginal distributions of the initial estimate of $\pi(\theta|s_{obs})$ are then adjusted to be those of the more precisely estimated marginals, through an appropriate replacement of order statistics. The final adjusted posterior can be a substantial improvement over standard ABC and regression adjustment methods  \shortcite{nott+fms14}.
While the marginal adjustment in itself avoids the curse of dimensionality problem, and can be applied to analyses with non-factorisable likelihood functions, the dependence structure within the initial estimate of $\pi(\theta|s_{obs})$ is not adjusted, and so the final marginally adjusted sample can have a very poor dependence structure.

In this article we propose a new method for constructing an ABC approximation to the posterior distribution that can be easily implemented in high dimensions, well beyond current ABC practice, while maintaining a viable dependence structure.
Our approach is based on constructing a Gaussian copula to approximate the dependence structure of $\pi(\theta|s_{obs})$, and on using the ideas behind the marginal adjustment to maintain full flexibility in representing the univariate margins. 
The
$p$-dimensional dependence structure of the Gaussian copula can be efficiently determined from the Gaussian copula dependence structures estimated from all bivariate parameter pairs $(\theta_i,\theta_j)$. As such, an advantage of this approach is that it plays to existing ABC method strengths: namely in only estimating low-dimensional (bivariate and univariate) posterior distributions. The copula approach accordingly overcomes the curse of dimensionality inherent in standard ABC methods, permitting the estimation of posterior distributions with viable dependence structures, for arbitrarily large $p$-dimensional parameter vectors.

This article is structured as follows:  In Section \ref{sec:methods} we introduce the Gaussian copula, and describe our proposed ABC method in detail. A simulated example and two real data analyses are presented in Section \ref{sec:examples}. The first real data analysis, based on the multivariate $g$-and-$k$ distribution, estimates a $p=184$ dimensional posterior distribution, which is, in principle, comfortably beyond the capabilities of any previous ABC analysis. Higher dimensional analyses could have been considered. 
The second real data analysis focuses on robust Bayesian variable selection, and illustrates how copula ABC can outperform both standard ABC and regular exact Bayesian inference even in moderate-dimensional analysis  (here $p=17$) in a discrete posterior setting.  Section \ref{sec:discussion} concludes with a discussion.

\section{Gaussian copula ABC}
\label{sec:methods}

According to the classical Bernstein-von Mises theorem (\shortciteNP{vandervarrt00}), under standard regularity conditions, the posterior distribution $\pi(\theta|y_{obs})$ is asymptotically normal. 
This motivates the use of structured density estimation models for ABC which contain the multivariate normal. In particular, we consider the meta-Gaussian family of distributions (\shortciteNP{fang+fk02}), which model dependence through a Gaussian copula, as we describe further below.  Meta-Gaussian densities 
have the property that the $p$-dimensional joint density can be reconstructed from all bivariate marginal densities. 
In the present setting, if bivariate marginal posterior densities can be well estimated using low-dimensional ABC analyses, then meta-Gaussian approximations to these densities can be combined into a meta-Gaussian approximation of the full posterior distribution. As it is constructed from well estimated marginal densities, the resulting posterior approximation would avoid the ABC curse of dimensionality problem, and can be expected to perform favourably compared to existing ABC approaches in high dimensional models.

Suppose that the random vector $\theta=(\theta_{1},\ldots,\theta_{p})^\top$ has a continuous multivariate density $g(\cdot)$, with univariate marginal densities $g_i(\cdot)$ and marginal distribution functions $G_i(\cdot)$ for $\theta_{i}$, $i=1,\ldots,p$. The copula $C$ of $\theta$ is defined as the joint distribution of $U=(U_{1},\ldots,U_{p})^\top=(G_1(\theta_{1}),\ldots,G_p(\theta_{p}))^\top$, and it contains full information on the dependence structure among the components of $\theta$.
Sklar's theorem (\shortciteNP{skalar59}) states that the multivariate density can be written as $g(\theta)=C(G_1(\theta_{1}),\ldots,G_P(\theta_{p}))$, which permits  a decoupling of the modelling of the copula and the univariate marginal densities in order to model the joint density (e.g. \shortciteNP{joe97}).

Define $\eta=(\eta_{1},\ldots,\eta_{p})^\top$ with $\eta_{i}=\Phi^{-1}(G_i(\theta_{i}))$, for $i=1,\ldots,p$, where $\Phi$ is the standard normal cumulative distribution function. 
If $\eta$ is multivariate normal, $\eta \sim N(0,\Lambda)$, then the copula of $\theta$ is called a Gaussian copula, and $\theta$ has a meta-Gaussian distribution with density function given by
\begin{eqnarray}
\label{copula}
	g(\theta)=\frac{1}{|\Lambda|^{1/2}}\mbox{exp}\left\{\frac{1}{2}\eta^\top (I-\Lambda ^{-1})\eta\right\}\prod_{i=1}^{p} g_i(\theta_{i}),
\end{eqnarray}
where $I$ denotes the identity matrix.

The multivariate normal family is embedded within the family of meta-Gaussian distributions. Writing $\phi(\cdot)$ as the standard normal density function, then the univariate normal distribution $N(\mu_1,\sigma^2_1)$ has density function $f(x_1)=\frac{\phi(\omega_1)}{\sigma_1}$ where $\omega_1=\frac{x_1-\mu_1}{\sigma_1}$. 
For a $p$-dimensional normal distribution $N(\mu,\Sigma)$, with mean $\mu=(\mu_1,\ldots,\mu_p)^\top$ and covariance matrix $\Sigma$, and writing $\omega=(\omega_{1},\cdots,\omega_{p})^\top$ with $\omega_{i}=\frac{x_i -\mu_i}{\sigma_i}$ for $i=1,\ldots,p$, then the joint normal density of $x=(x_1,\ldots,x_p)^\top$ can be expressed as
\begin{eqnarray}
	f(x)&=& \frac{1}{(2\pi )^{p/2}|\Sigma|^{1/2}}\mbox{exp}\left\{-\frac{1}{2}(x-\mu)^\top \Sigma^{-1}(x-\mu)\right\}    \nonumber \\
	&=& \frac{1}{(2\pi)^{p/2}|R|^{1/2}} \mbox{exp}\left\{ -\frac{1}{2}{\omega}^\top R^{-1}\omega \right\}{\prod_{i=1}^{p}\frac{1}{\sigma_i}} \nonumber \\
	&=& \frac{1}{|R|^{1/2}} \mbox{exp}\left\{ \frac{1}{2} {\omega}^\top (I-R^{-1})\omega \right\} \prod_{i=1}^{p}\frac{\phi(\omega_i)}{\sigma_i},\label{normal}
\end{eqnarray}
where $R$ is the corresponding correlation matrix of $\Sigma$. Observe that $R$ in (\ref{normal}) corresponds to $\Lambda$ in (\ref{copula}), meaning that the correlation matrix of the Gaussian distribution is exactly the correlation matrix of the corresponding Gaussian copula.

In the ABC setting, if approximate normality of $\pi(\theta|s_{obs})$ holds, possibly after marginal transformations of the parameters, then we may utilise a Gaussian copula model to estimate the dependence structure of
$\pi(\theta|s_{obs})$
in light of (\ref{normal}). 
As previously noted, all bivariate marginal densities completely determine the joint density in a meta-Gaussian distribution, and these bivariate densities can usually be easily and precisely estimated in low-dimensional ABC analyses. As such, it will be possible to obtain a reliable estimate of the joint posterior $\pi(\theta|s_{obs})$, even in high-dimensional problems, something which is in principle comfortably beyond the capabilities of current ABC methods. 
In essence, we propose to estimate each bivariate density using low-dimensional ABC methods, approximate these with a $2$-dimensional Gaussian copula, and then combine them to obtain an approximate joint posterior using (\ref{copula}).

More precisely, the procedure we propose  is as follows:
\begin{enumerate}
\item\label{step1} For each pair $(i,j)$ with $i=1, \cdots, p-1$ and $j=i+1,\ldots, p$:
\begin{enumerate}
\item 
Identify the summary statistics $s_{(i,j)}$ as a subset of $s$ which are informative for $(\theta_{i},\theta_{j})$. 
\item \label{step1b}Use conventional ABC methods to draw an approximate sample $\theta^{(1)},\ldots,\theta^{(n)}$ from $\pi(\theta|s_{(i,j)}$). Extract the $(i,j)^{th}$ components from $\theta^{(1)},\ldots,\theta^{(n)}$ to form an approximate sample $(\theta^{(1)}_i,\theta^{(1)}_j),\ldots,(\theta^{(n)}_i,\theta^{(n)}_j)$ from the bivariate marginal 
$\pi(\theta_i,\theta_j|s_{(i,j)})$.

\item
Let $r^{(1)}_i,\ldots,r^{(n)}_i$ be the ranks of  $\theta^{(1)}_i,\ldots,\theta_i^{(n)}$, and $q^{(1)}_j,\ldots,q^{(n)}_j$ be the ranks of  $\theta^{(1)}_j,\ldots,\theta_j^{(n)}$. 
Set $\eta_i^{(\ell)}=\Phi^{-1}(\frac{r^{(\ell)}_i }{n+1})$ and $\eta_j^{(\ell)}=\Phi^{-1}(\frac{q^{(\ell)}_j }{n+1})$ for  $\ell=1,\ldots, n.$
\item
Calculate the sample correlation of $(\eta_{i}^{(1)},\eta_{j}^{(1)}),\ldots,(\eta_i^{(n)},\eta_j^{(n)})$ and denote it $\hat{\Lambda }_{i,j}$ ($=\hat{\Lambda}_{j,i}$).
\end{enumerate}

\item \label{step2} For $i=1,\ldots,p$:
\begin{enumerate}
\item Identify the summary statistics $s_{(i)}$ as a subset of $s$ which are informative for $\theta_i$.
\item Use conventional ABC methods to draw an approximate sample $\theta^{(1)},\ldots,\theta^{(n')}$ from $\pi(\theta|s_{(i)})$. Extract the $i^{th}$ component from $\theta^{(1)},\ldots,\theta^{(n')}$ to form an approximate sample $\theta_i^{(1)},\ldots,\theta_i^{(n')}$ from the univariate marginal $\pi(\theta_i|s_{(i)})$.
\item Use density estimation methods to approximate the marginal density $g_i(\theta_i)$ (denoted $\hat{g}_i(\theta_i)$) based on $\theta^{(1)}_i,\ldots,\theta^{(n')}$.
\end{enumerate}

\item Combine all $\hat{\Lambda} _{i,j}$'s to form the $p$-dimensional correlation matrix $\hat{\Lambda}$ with diagonal elements $1$. The final Gaussian copula estimate of $\pi(\theta|s_{obs})$ is obtained via $(\ref{copula})$ with $\Lambda$ estimated by $\hat{\Lambda}$ and $g_i(\theta_i)$ estimated by $\hat{g}_i(\theta_i)$ for $i=1,\ldots,p$.

\end{enumerate}

The above algorithm is easy to implement, and is computationally efficient as the calculations in Steps \ref{step1} and \ref{step2} can be performed in parallel for each $i,j$. While there is no restriction on the types of ABC methods used to draw approximate samples from $\pi(\theta_i,\theta_j|s_{(i,j)})$ and $\pi(\theta_i|s_{(i)})$, one possible efficient implementation could be to estimate all bivariate and univariate marginal densities using importance sampling (Table \ref{eqn:importanceSampling}) using the same large initial sample $(\theta^{(\ell)},s^{(\ell)})\sim L(s|\theta)p(\theta)$ for $\ell=1,\ldots,N$. This approach is common in the ABC literature (e.g. \shortciteNP{nunes+b10,blum+nps13,prangle+bps14}), and is one we adopt in the analyses of Section \ref{sec:examples}. 
Alternatively, separate samplers could be implemented (in parallel) for each univariate and bivariate margin, although at potentially higher computational overheads.

A key element in the accurate estimation of the bivariate and univariate marginal densities is the identification of suitable subsets of statistics $s_{(i,j)}$ and $s_{(i)}$. While this may initially seem difficult, it is not uncommon to be able to identify specific summary statistics as informative for specific parameters, particularly in some structured models (e.g. \shortciteNP{drovandi+p11,nott+fms14}). However, in more general cases, established techniques exist for the semi-automatic construction of a single summary statistic for each model parameter \cite{fearnhead+p12}. This method is particularly useful in the present framework.

As the meta-Gaussian distribution (\ref{copula}) is used as an approximation to $\pi(\theta|s_{obs})$, it is sensible to examine the quality of the final approximation. This can be achieved through existing diagnostic procedures for ABC approximations \shortcite{prangle+bps14}, or during the construction of the copula model itself. For the latter, note that bivariate Gaussian copula models for each $g(\theta_i,\theta_j)$ are available through (\ref{copula}), and can be estimated as $\tilde{g}_{ij}(\theta_i,\theta_j)$ given $\hat{g}_i(\theta_i)$, $\hat{g}_j(\theta_j)$ and $\hat{\Lambda}_{i,j}$. Similarly, a bivariate kernel density estimate of $\pi(\theta_i,\theta_j|s_{obs})$, denoted $\hat{g}_{ij}(\theta_i,\theta_j)$, can be constructed from the samples $(\theta^{(1)}_i,\theta^{(1)}_j),\ldots,(\theta^{(n)}_i,\theta^{(n)}_j)$ in Step \ref{step1b}.

If approximate normality of the posterior holds, then the bivariate dependence structure can be well described by a Gaussian copula, and hence $\tilde{g}_{ij}(\theta_i,\theta_j)$ will provide a close approximation to $\hat{g}_{ij}(\theta_i,\theta_j)$.
If for every bivariate pair $(\theta_i,\theta_j)$ the Gaussian copula estimate $\tilde{g}_{ij}(\theta_i,\theta_j)$ provides a close approximation to $\hat{g}_{ij}(\theta_i,\theta_j)$, this suggests that 
the full posterior may be adequately modelled by a Gaussian copula. 
Of course, capturing all bivariate dependence structures well does not necessarily mean that the full joint dependence will be captured well. 
As such, some kind of application specific predictive validation of the approximate joint posterior may be needed.

Finally, we note that   
the estimate $\hat{\Lambda}$ obtained by combining the $\hat{\Lambda}_{i,j}$ is not guaranteed to be positive definite (although in all our later analyses it was).  If this occurs then alternative procedures for constructing $\hat{\Lambda}$ can be adopted, such as the methods considered in \shortciteN{loland+hhf13}.  We also note that the use of a plug-in estimator for $\Lambda$ ignores the possibility of large estimation errors. If this is a realistic possibility in any analysis, then a sensitivity analysis should be performed.

\section{Examples}
\label{sec:examples}

\subsection{A toy example}
\label{section:toy}

We first examine how the ABC Gaussian copula posterior (\ref{copula}) performs in a simple toy example, where the posterior distribution is known. The model that we consider is $y\sim N_p(\theta,\Sigma)$ for $p\geq 2$, where $y=(y_1,\ldots,y_p)^\top$, $\theta=(\theta_1,\ldots,\theta_p)^\top$ and $\Sigma=\mbox{diag}(\sigma_0,\ldots,\sigma_0)$. For the prior we specify the `twisted-normal' prior of \shortciteN{haario+st99} with density function proportional to
\[
	p(\theta)\propto \exp\left\{-\frac{\theta_1^2}{200} - \frac{(\theta_2-b\theta_1^2+100b)^2}{2} - \sum_{j=3}^p\theta_j^2\right\}.
\]
For $p=2$, the third term in the exponent is set to be zero.
This prior is essentially a product of independent Gaussian distributions with the exception that the component for $(\theta_1,\theta_2)$ 	is modified to produce a `banana' shape, with the strength of the bivariate dependence determined by the parameter $b$.  Simulation from $p(\theta)$ is achieved by first drawing $\theta\sim N_p(0,A)$ where $A=\mbox{diag}(100,1,\ldots,1)$ and then transforming $\theta_2\rightarrow\theta_2+b\theta_1^2-100b$.

For the following we specify $\sigma_0=1$, and $b=0.1$ to produce strong prior dependence between $\theta_1$ and $\theta_2$.
We determine $y_{obs}=(10,0,\ldots,0)^\top$ as a single observed vector, and construct the vector of summary statistics as $s=S(y)=y$, the full, $p$-dimensional dataset. We exploit knowledge of the model and set $s_{(i)}=s_i$ as the subset of summary statistics that are informative for $\theta_i$, with the exception of $s_{(2)}=(s_1,s_2)$ for $\theta_2$. The unions of these informative subsets $s_{(i)}$ and $s_{(j)}$ are taken when constructing the subsets $s_{(i,j)}$ informative for the bivariate parameter pair $(\theta_i,\theta_j)$.

The following analyses are based on $N=1,000,000$ samples $(\theta^{(\ell)},s^{(\ell)})\sim L(s|\theta)\pi(\theta)$, $\ell=1,\ldots,N$.
In sampling from each 1- and 2-dimensional ABC posterior approximation, as required to construct the Gaussian copula approximation $\tilde{g}(\theta)$ of $g(\theta)$, we specify the smoothing kernel $K_h(\cdot)$ as uniform over the range $(-h, h)$ and determine $h$ as the 0.01 quantile of the $N$ observed differences between simulated and observed summary statistics (with different summary statistics for each marginal-posterior approximation), producing $n=n'=10,000$ equally weighted samples for analysis. In each case, both local linear regression-adjustment \shortcite{beaumont+zb02} and marginal adjustment \shortcite{nott+fms14} were implemented to improve the posterior approximation.
Euclidean distance $\|s-s_{obs}\|=[\sum_{i=1}^p(s_i-s_{obs,i})^2]^{1/2}$ was used to compare simulated and observed summary statistics.

\begin{figure}[tbh]
\centering
\includegraphics[width=15.5cm]{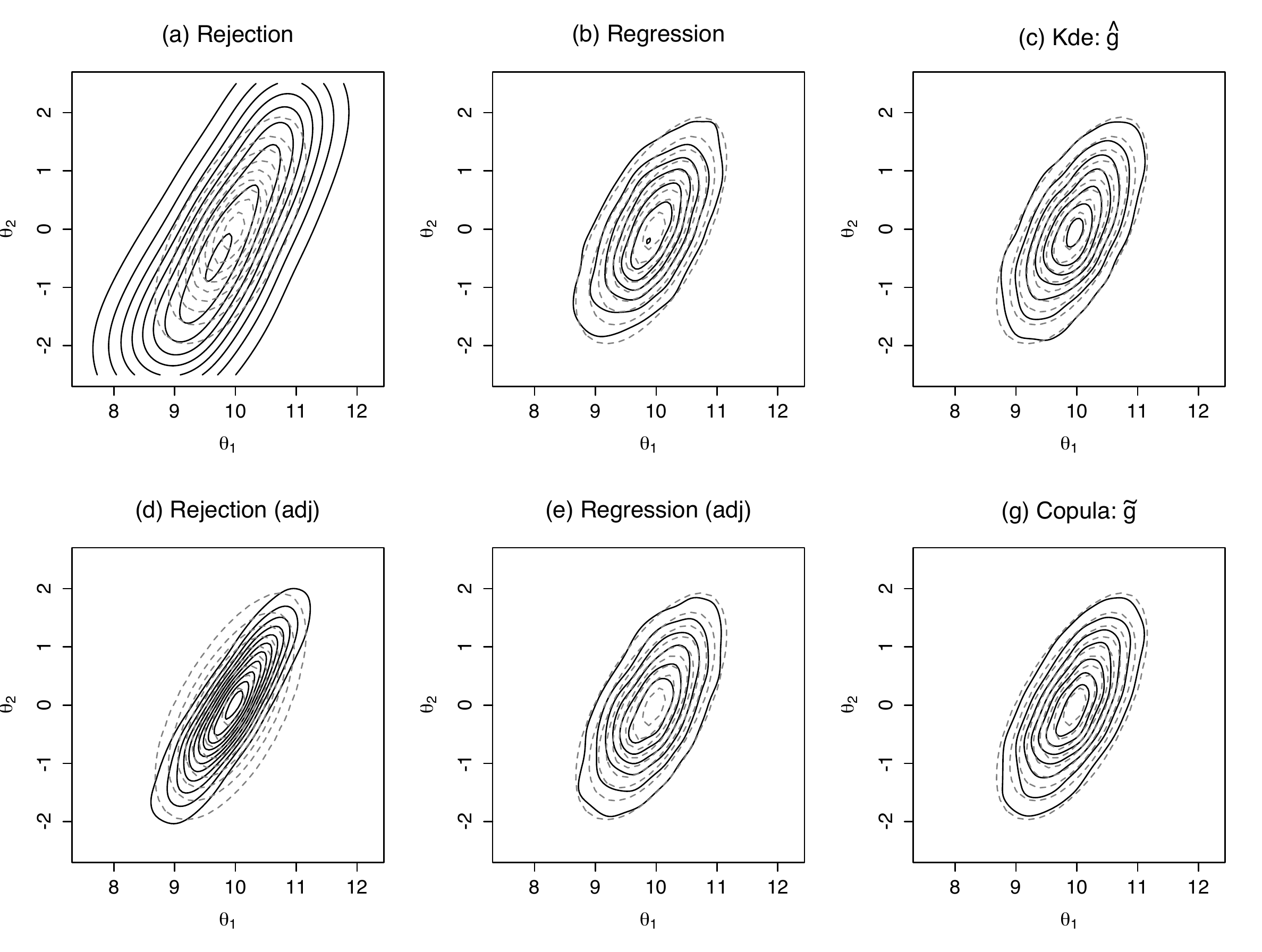}
\caption{\small Contour plots  of the $(\theta_1,\theta_2)$ margin of various ABC posterior approximations (black lines) to the $p=5$ dimensional model, $\pi(\theta|s_{obs})$. True contours are shown in grey-dashed lines, and contour levels indicate 0.1, \ldots, 0.9 of maximum density estimate. Standard ABC approximations consist of (a) rejection sampling, (b) rejection sampling with regression adjustment, (d) rejection sampling with marginal adjustment, and (e) rejection sampling, with regression and marginal adjustment. Panel (c) illustrates regression and marginal adjusted estimate $\hat{g}_{1,2}(\theta_1,\theta_2)$ of $\pi(\theta_1,\theta_2|s_{(1,2)})$, whereas panel (g) shows the copula ABC approximation $\tilde{g}_{1,2}(\theta_1,\theta_2)$.}
\label{fig:banana5D}
\end{figure}

Figure \ref{fig:banana5D} illustrates contour plots of various estimates of the bivariate posterior margin $\pi(\theta_1,\theta_2|s_{obs})$ (solid lines), each derived from estimates of the full distribution $\pi(\theta|s_{obs})$ when $p=5$. Contour plots of the true bivariate margin are given by the grey dashed lines.
The left column of Figure \ref{fig:banana5D} shows the estimates obtained via standard rejection ABC using the full vector of summary statistics $s_{obs}$, both without (panel (a)) and with (panel (d)) marginal adjustment. The univariate margins for the marginal adjustment were obtained from the $p=2$ dimensional model. 
From panel (a), rejection sampling alone performs fairly poorly -- the correlation between $\theta_1$ and $\theta_2$ is captured reasonably well, but the univariate margins are too dispersed. Following a marginal adjustment (panel (d)), the margins are corrected to the right scale, but now it becomes evident that the dependence structure is not perfectly estimated.

The centre column of Figure \ref{fig:banana5D}, shows the same information as the rejection-based estimates, except that a linear regression adjustment has been performed in each case after the rejection stage, and before the marginal adjustment. Clearly the regression adjusted samples (panel (b)) approximate the true posterior very well, to the extent that no further visual improvements are apparent following a subsequent marginal adjustment (panel (e)).

Panel (c) displays the kernel density estimates $\hat{g}_{1,2}(\theta_1,\theta_2)$, obtained following regression and marginal adjustments, but where each margin is only conditioned on the subvector of summary statistics $s_{(1,2)}$ rather than on the full vector $s_{obs}$. That the kernel density estimates are largely the same as for the standard ABC analyses indicates that the subvector $s_{(1,2)}$ is highly informative for the bivariate parameter pair, and that these are therefore appropriate to use when fitting the copula model.

Panel (g) shows the fitted bivariate copula estimates $\tilde{g}_{1,2}(\theta_1,\theta_2)$ based on (\ref{copula}). As the copula ABC approximation is highly similar to the kernel density estimate $\hat{g}_{1,2}(\theta_1,\theta_2)$, this indicates that the copula model is both appropriate and accurate for these bivariate margins. 
Similar qualitative comparisons can be made for all other bivariate marginal distributions (results not shown), implying that the full copula model $g(\theta)$ may be extended as a good approximation of $\pi(\theta|s_{obs})$.

\begin{figure}[tbh]
\centering
\includegraphics[width=15.5cm]{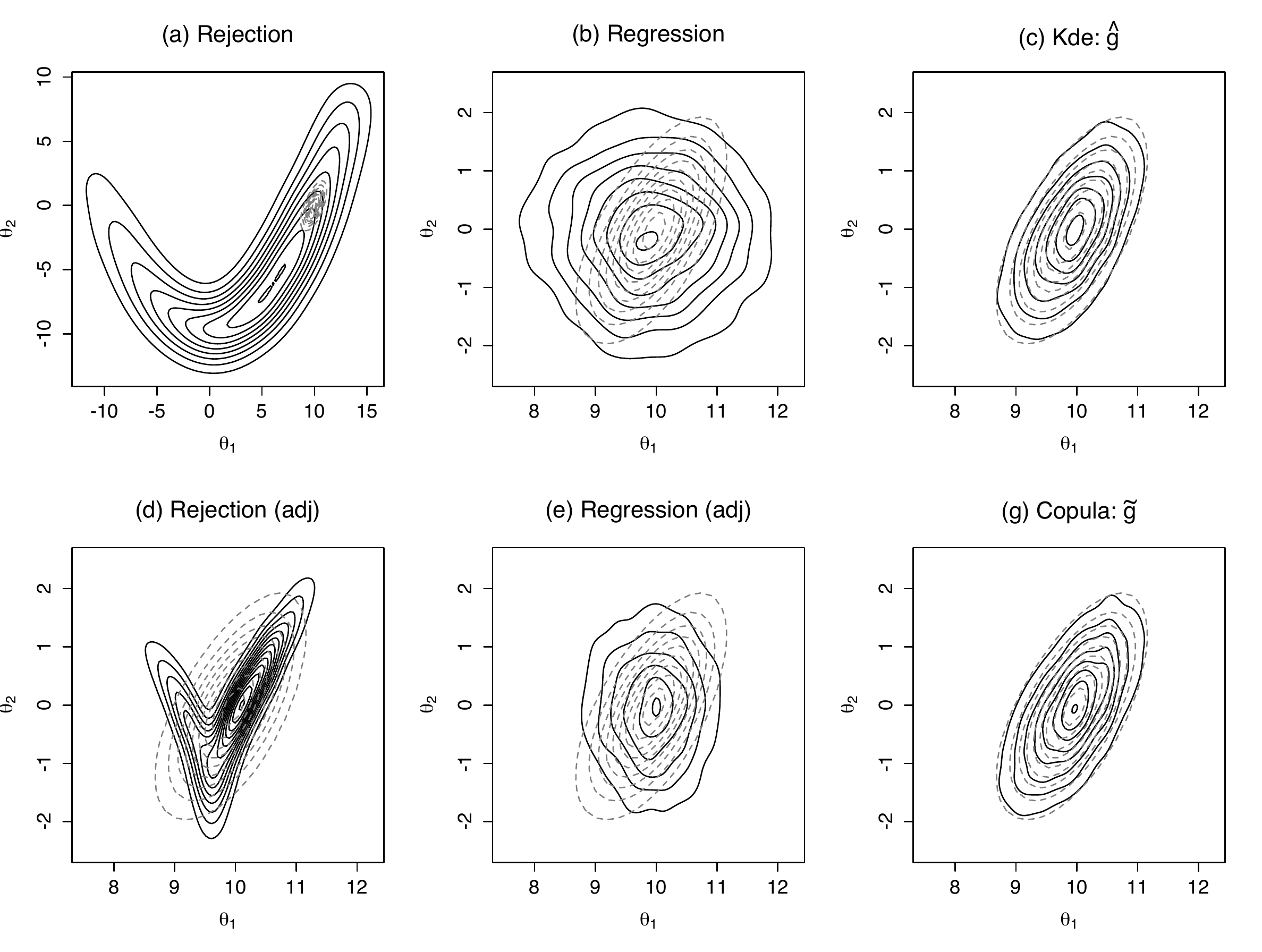}
\caption{\small Contour plots  of the $(\theta_1,\theta_2)$ margin of various ABC posterior approximations (black lines) to the $p=50$ dimensional model, $\pi(\theta|s_{obs})$. True contours are shown in grey-dashed lines and contour levels indicate 0.1, \ldots, 0.9 of maximum density estimate. Standard ABC approximations consist of (a) rejection sampling, (b) rejection sampling with regression adjustment, (d) rejection sampling with marginal adjustment, and (e) rejection sampling, with regression and marginal adjustment. Panel (c) illustrates regression and marginal adjusted estimate $\hat{g}_{1,2}(\theta_1,\theta_2)$ of $\pi(\theta_1,\theta_2|s_{(1,2)})$, whereas panel (g) shows the copula ABC approximation $\tilde{g}_{1,2}(\theta_1,\theta_2)$.}
\label{fig:banana50D}
\end{figure}

Figure \ref{fig:banana50D} shows the same estimates of the bivariate margin $\pi(\theta_1,\theta_2|s_{obs})$ as Figure \ref{fig:banana5D}, except that they are derived from estimates of the full distribution $\pi(\theta|s_{obs})$ when $p=50$. In this scenario, the limitations of standard ABC methods become apparent. Due to the increased number of parameters, $p$, the rejection sampling estimate of the margin $\pi(\theta_1,\theta_2|s_{obs})$ is highly similar to the `banana' prior distribution, $\pi(\theta)$. This deviation cannot be corrected by adjusting the margins (panel (d)). The regression adjusted estimate (panel (b)) performs better -- it is centered on the right location, although the margins are too diffuse, and the posterior correlation has disappeared. Correcting the margins (panel (e)) improves this aspect, although it cannot recover the lost dependence structure.
In comparison, the copula marginal estimate $\tilde{g}_{1,2}(\theta_1,\theta_2)$ retains the same accuracy as for the $p=5$ dimensional model as it is constructed in exactly the same way.

\begin{table}[tbh]
\centering
\begin{tabular}{llllll}
  & Rejection & {\centering Rejection} & Regression & Regression & Copula ABC \\
 $p$ & & (Marginal adj.) & & (Marginal adj.)  \\
\hline
2      & 0.058 ($<$0.001)& 0.040 ($<$0.001) & 0.043 ($<$0.001) & 0.035 ($<$0.001) & 0.039 ($<$0.001)\\
5      & 0.807 (0.001)       & 0.053 (0.001)        & 0.613 (0.002)        & 0.037 ($<$0.001) & 0.040 ($<$0.001)\\
10    & 1.418 (0.002)       & 0.100 (0.001)        & 1.078 (0.002)        & 0.061 (0.001)        & 0.040 ($<$0.001)\\
15    & 1.912 (0.002)       & 0.292 (0.002)        & 1.229 (0.003)        & 0.202 (0.001)        & 0.039 ($<$0.001)\\
20    & 2.288 (0.002)       & 0.450 (0.001)        & 1.280 (0.003)        & 0.292 (0.001)        & 0.039 ($<$0.001)\\
50    & 3.036 (0.003)       & 0.520 (0.002)        & 1.474 (0.009)        & 0.335 (0.001)        & 0.040 ($<$0.001)\\
100  & 3.362 (0.002)       & 0.524 (0.002)        & 1.619 (0.013)        & 0.341 (0.001)        & 0.039 ($<$0.001)\\
250  & 3.663 (0.003)       & 0.515 (0.002)        & 1.737 (0.015)        & 0.344 (0.001)        & 0.039 ($<$0.001)\\
\hline
\end{tabular}
\caption{\small Estimated Kullback-Leibler divergence of the $(\theta_1,\theta_2)$ margin of various ABC posterior approximations to $\pi(\theta_1,\theta_2|s_{obs})$, as a function of model dimension $p$. Numbers represent mean divergences over 100 replicates with standard errors given in parentheses.
\label{table:toy}}
\end{table}

To illustrate more precisely the performance of each ABC posterior estimation method as dimension $p$ increases, Table \ref{table:toy} shows the mean estimated Kullback-Leibler (KL) divergence between $\pi(\theta_1,\theta_2|s_{obs})$ and the bivariate margin of each ABC approximation, based on 100 replicates. The number in parentheses is the standard error of this estimate. As dimension increases, the performance of rejection ABC deteriorates drastically, as expected. As $p$ gets very large, the KL divergence will level off to that obtained by comparing $\pi(\theta_1,\theta_2|s_{obs})$ to the bivariate `banana' prior $p(\theta_1,\theta_2)$, as the ABC estimate of the posterior becomes equivalent to that prior as $p\rightarrow\infty$.  The marginally adjusted rejection sample performs better, though only because it at worst maps the `banana' prior to the region of high posterior density -- it otherwise performs poorly (see e.g. Figure \ref{fig:banana50D}(d)).

The regression-adjusted estimates perform better than the rejection ABC estimates, as they exploit the linear relationship between $\theta_i$ and $s_i$ in order to better identify the high posterior density region. However, even regression adjustment is known to only mitigate the curse of dimensionality in ABC \shortcite{nott+fms14}. As $p$ gets large, the best performance will be obtained by performing regression adjustment on the prior distribution (which is the limiting approximation for rejection sampling). Performing the marginal adjustment can improve on this, but as all dependence structure has been lost with higher dimensions, the best possible approximation here is a product of the independent marginal estimates \shortcite{nott+fms14}.

In contrast, the copula ABC approach is constructed from low dimensional (i.e. bivariate) estimates of $\pi(\theta_i,\theta_j|s_{obs})$, regardless of the dimension of the full model. As such, it can near perfectly capture the dependence structure of all bivariate pairs of the full posterior distribution, which is near Gaussian in this example. That is, its performance is completely independent of model dimension.

\subsection{A high-dimensional, multivariate g-and-k model}
\label{section:gandk}

A multivariate version of the $g$-and-$k$ distribution was introduced by \shortciteN{drovandi+p11}. This $q$-dimensional distribution is constructed by linking $q$ univariate $g$-and-$k$ marginal distributions (\shortciteNP{raynerm02}), with marginal parameters $(A_i,B_i,g_i,k_i)$ for $i=1,\ldots,q$, together with a Gaussian copula with correlation matrix $V$ for the dependence structure. 
The univariate $g$-and-$k$ distribution has no closed form density, but is defined through its quantile function as
\begin{eqnarray}
	\label{gk}
	Q(q|A,B,g,k)=A+B\left[1+c\frac{1-\mbox{exp}\{-g z(q)\}}{1+\mbox{exp}\{-g z(q)\}}\right](1+z(q)^2)^k z(q),
\end{eqnarray}
for $B>0,\;k>-1/2$,
where the parameters $A,B,g$ and $k$ control location, scale, skewness and kurtosis respectively, and where $z(q)$ denotes the $q$-th quantile of the standard normal distribution function.   The parameter $c$ measures the overall asymmetry, and is fixed at $0.8$ as a conventional choice (\shortciteNP{raynerm02}).  Several ABC approaches to inference for the univariate $g$-and-$k$ and related distributions have previously been considered \shortcite{allinghamkm09,drovandi+p11,fearnhead+p12,peters+s06}.  The univariate $g$-and-$k$ distribution is very flexible, with many common distributions obtained or well approximated by appropriate parameter settings, such as the normal distribution when $g=k=0$.
Given $(A,B,g,k)$, simulations $z(p)\sim N(0,1)$ drawn from a standard normal distribution can be transformed into samples from the $g$-and-$k$ distribution through (\ref{gk}). To obtain draws from the multivariate model, first draw samples from $N_q(0,V)$, and then adjust each of the $q$ margins as for the univariate case. 

Note that the use of a Gaussian copula for the multivariate $g$-and-$k$ distribution is completely distinct from our use of a Gaussian copula to approximate $\pi(\theta|s_{obs})$ through (\ref{copula}). However, the copula construction of the multivariate $g$-and-$k$ distribution does permit the ABC analysis of  a single model type with an arbitrarily large number of parameters. The number of unknown parameters in this model consists of the four parameters $(A_i,B_i,g_i,k_i)$ for each of the $q$ univariate margins, plus $q(q-1)/2$ correlation parameters $\nu_{ij}=\nu_{ji}$ for $i,j=1,\ldots,q$, in the correlation matrix $V=[\nu]_{ij}$ of the $g$-and-$k$ copula. This gives $4q+q(q-1)/2$ parameters in total for the $q$-dimensional model.

The observed data consist of $q=16$ foreign currency exchange log daily returns against the Australian dollar (AUD) for 1,757 trading days from 1st January 2007 to 31st December 2013 \cite{rba14}.  Hence, our most complex model has 184  unknown parameters. This is considerably beyond the scope of any previous ABC analysis that does not rely on likelihood factorisation to perform the analysis.

For the univariate model margins, \shortciteN{drovandi+p11}  proposed the following robust summary statistics as informative for the four model parameters:
\begin{eqnarray*}
	S_A=L_2, &  & S_k=(E_7-E_5+E_3-E_1)/S_B,\\
	S_B=L_3-L_1, &  \mbox{and} & S_g=(L_3+L_1-2L_2)/S_B,
\end{eqnarray*}
where $L_i$ and $E_j$ respectively denote the $i$-th sample quartile and $j$-th octile of the dataset $y$.  We adopt these statistics as directly informative for each respective model parameter in defining $s_{(i)}$, so that e.g. $S_A$ is informative for $A$ and $S_g$ is informative for $g$. The exception to this is that we specify $(S_B,S_k)$ as informative for $B$. The dependence of $B$ on both of these statistics is immediately apparent by regressing $\theta$ on $s=(S_A,S_B,S_g,S_k)$ using the $N$ samples $(\theta^{(\ell)},s^{(\ell)})$, in the mould of \citeN{fearnhead+p12}. 
Also following \shortciteN{drovandi+p11}, we use the robust normal scores correlation coefficient \cite{fisher+y48} as the informative summary statistic for each correlation parameter $\nu_{ij}$ between the $i$-th and $j$-th data margins.
The unions of these informative subsets $s_{(i)}$ and $s_{(j)}$ are taken when constructing the subsets $s_{(i,j)}$ informative for bivariate parameter pairs. So e.g. $(S_g,S_k)$ and $(S_A,S_B,S_k)$ are taken as informative for $(g,k)$ and $(A,B)$ respectively.
The prior $p(\theta)$ is defined as uniform over the support of the parameter space for the $(A_i,B_i,g_i,k_i)$ margins and a $\mbox{Wishart}(I_q,q)$ distribution with $q$ degrees of freedom for $V$, where $I_q$ denotes the $q\times q$ identity matrix.

The following analyses are based on $N=500,000$ samples $(\theta^{(\ell)},s^{(\ell)})\sim L(s|\theta)f(\theta)$, $\ell=1,\ldots,N$, where the importance sampling distribution $f(\theta)$ is defined by $U(-0.1,0.1)\times U(0,0.05)\times U(-1,1)\times U(-0.2,0.5)$ for each $g$-and-$k$ marginal parameter set $(A_i,B_i,g_i,k_i)$, and the $\mbox{Wishart}(I_q,q)$ prior distribution for the correlation matrix $V$. The uniform range for each marginal parameter was determined via a pilot analysis using a moderate number of samples $(\theta^{(\ell)},s^{(\ell)})$, following \citeN{fearnhead+p12}. The smoothing kernel $K_h(\cdot)$ is uniform over $(-h,h)$ where $h$ is determined as the 0.01 quantile of the $N$ differences between simulated and observed summary statistics $\|s^{(\ell)}-s_{obs}\|$. Mahalanobis distance was used to compare simulated and observed summary statistics $\|s-s_{obs}\|=[(s-s_{obs})'\Sigma_0^{-1}(s-s_{obs})]^{1/2}$, where $\Sigma_0=\mbox{Cov}(s|\theta_0)$ was estimated as the sample covariance of 2000 samples from $L(s|\theta_0)$, and where $\theta_0$ is determined as the vector of means of the marginal density estimates $\hat{g}_i(\theta_i)$ $i=1,\ldots,p$.

\begin{figure}[tbh]
\centering
\includegraphics[width=15.5cm]{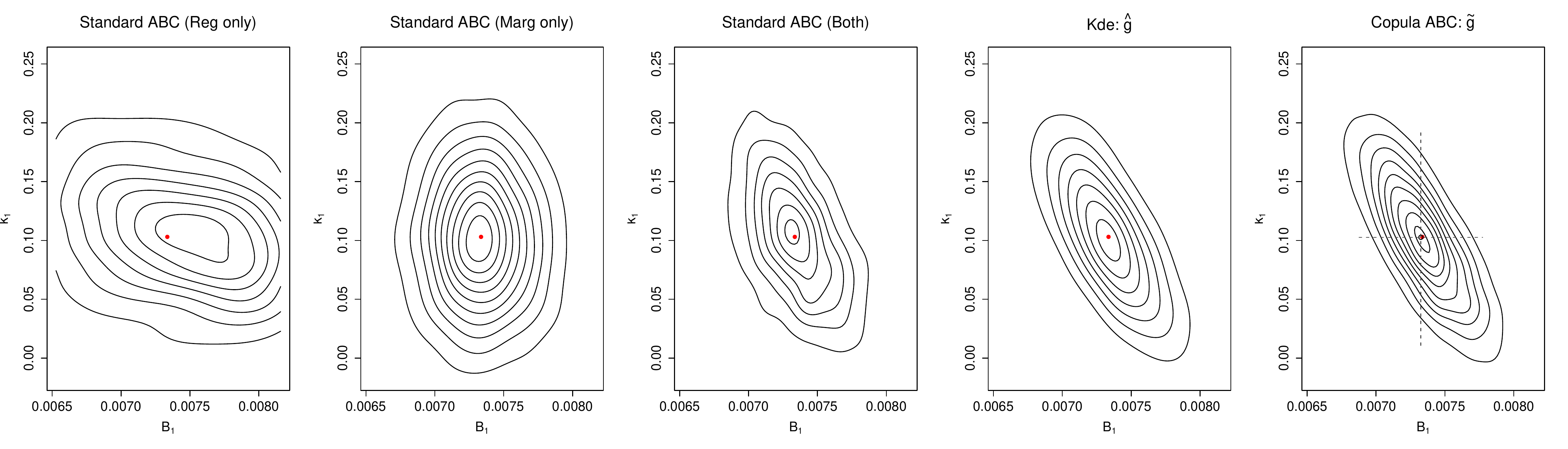}
\includegraphics[width=15.5cm]{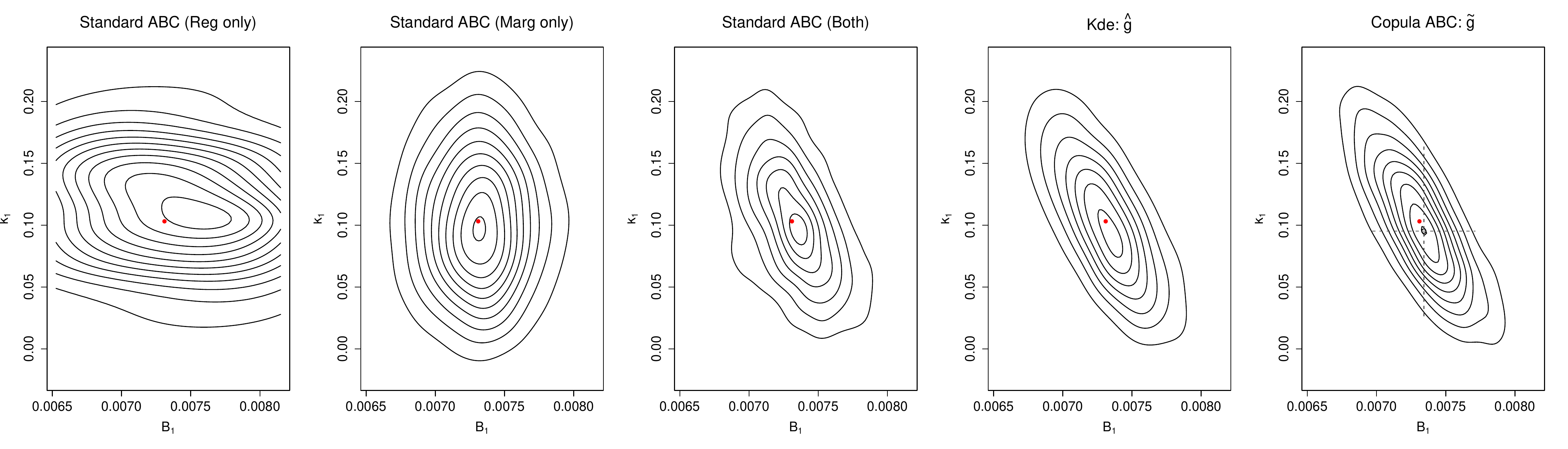}
\includegraphics[width=15.5cm]{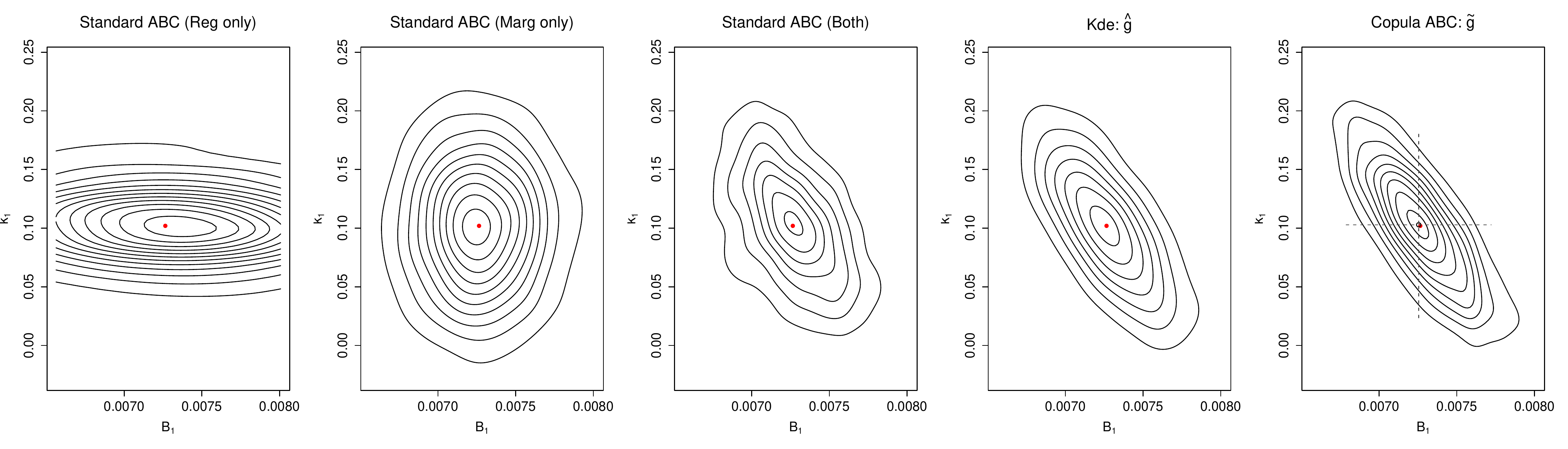}
\caption{\small Contour plots  of the $(B_1,k_1)$ margin of various ABC posterior approximations to the multivariate $g$-and-$k$ model, $\pi(\theta|s_{obs})$. 
Rows correspond to the $q=3$ (top), $q=10$ and $q=16$ (bottom) dimensional model, which have $p=15$, $p=85$ and $p=184$ parameters respectively.
Standard ABC approximations consist of (column 1) rejection sampling with regression adjustment, (column 2) rejection sampling with marginal adjustment, and (column 3) rejection sampling, with regression and marginal adjustment. Column 4 illustrates the regression and marginal adjusted kernel density estimate $\hat{g}(B_1,k_1)$ of $\pi(\theta_i,\theta_j|s_{(i,j)})$, whereas column 5 shows the corresponding regression and marginal adjusted copula ABC approximation $\tilde{g}(B_1,k_1)$. The dot in each panel indicates the value of $\theta_0$ used to estimate $\Sigma_0$ in the Mahalanobis distance calculation. The crosshairs in column 5 show the marginal MLE plus or minus approximately two posterior standard deviations.}
\label{fig:gk}
\end{figure}

Figure \ref{fig:gk} illustrates contour plots of various ABC approximations of the bivariate $(B_1,k_1)$ posterior marginal distribution. The top row corresponds to the $q=3$-dimensional model with $p=15$ parameters. The middle and bottom rows correspond to $q=$ 10- and 16-dimensional models with $p=$ 85 and 184 parameters respectively.  
Column 1 shows the ABC posterior approximation based on importance sampling and regression adjustment only. Clearly the approximation is poor, regardless of the dimension of the model, with the approximation becoming slightly variable as model dimension increases. The density estimates in column 2 are based on rejection sampling and marginal adjustment only. Here, while the marginal adjustment brings the posterior approximation to the right scale, the previously observed negative dependence structure between $B_1$ and $k_1$ has been lost due to the initially poor rejection sampling estimate (not shown).

Column 3 of Figure \ref{fig:gk} illustrates conventional best performance ABC: rejection sampling followed by both regression and marginal adjustments. In this scenario the estimated marginal posterior seems credible for any model dimension, displaying viable scale and negative dependence structure. Column 4 shows the same posterior approximation as column 3, except that only the subset of summary statistics $s_{(i,j)}=(S_{B_1},S_{k_1})$ is used in the estimation, rather than the full $p$-dimensional vector, $s$. As the density estimate is broadly equivalent to that using the full vector of summary statistics, this indicates that the subset $s_{(i,j)}$ is indeed highly informative for this parameter pair. Moreover,  the estimate of $\pi(B_1,k_1|s_{(i,j)})$ is more precisely estimated than the estimate of $\pi(B_1,k_1|s)$, indicating that there is some effect on the standard ABC approximation as the dimension of the vector summary statistics gets large. This loss of precision is not seen when only using the summary statistic subset $s_{(i,j)}.$
Finally, column 5 displays the bivariate copula margin estimate of $\pi(B_1,k_1|s)$, which is effectively the same as the kernel density estimate in column 4. This indicates that the copula model provides a good approximation for this bivariate marginal distribution.

What is notable in this analysis is that standard ABC methods are performing admirably well, even in $p=184$ dimensions. Chiefly this is due to the relationships between the sampled summary statistics and parameter pairs $(\theta_i^{(\ell)},s_{(i)}^{(\ell)})$ being highly linear, in combination with the structured construction of the multivariate $g$-and-$k$ model. The former point enables the regression adjustment to estimate the linear dependence structure between parameter pairs well, 
whereas the latter point means that the parameters $(A_i,B_i,g_i,k_i)$ of the $i$-th margin, are mostly (but not completely) determined by the data in the same margin.
In combination with the marginal adjustment, these allow standard ABC methods to produce very good estimates of the posterior distribution.  However, some improvement is still clearly
being brought by the copula approach.  These results imply that while the original paper that developed this multivariate quantile model for ABC only analysed data with $q=2$ dimensions ($p=9$ parameters) \cite{drovandi+p11}, this model is clearly viable for inference in much higher dimensions.

We use this example to illustrate another advantage of the copula approach.  The fitted copula model $\tilde{g}(\theta)$ provides an analytic approximation to the posterior distribution, $\pi(\theta|s_{obs})$.  
From Bayes' rule we have $L(s_{obs}|\theta)\propto \pi(\theta|s_{obs})/p(\theta)$, and 
hence $\tilde{g}(\theta|y_{obs})/p(\theta)$ is an approximation of a function proportional to the likelihood.  We can use this approximation to compute approximations of maximum likelihood estimates and the observed information matrix and hence perform frequentist analyses that can be used for comparison with the full Bayes analysis. It is also possible to compute marginal likelihoods for subsets of parameters after integrating out the other parameters according to the conditional prior. 
\citeN{grazian+l15} recently considered the use of ABC for this purpose based on kernel esitmation of the ABC marginal posterior.  If the parameter of interest
is of moderate dimension it may be difficult to implement kernel estimation however. For our copula method the idea is
illustrated in column 5 of Figure \ref{fig:gk}, where the open circles denote the approximate marginal MLE for $(B_1,k_1)$ -- obtained by maximising $\tilde{g}_{1,2}(B_1,k_1)/p(B_1,k_1)$ for each respective model --
and the dashed crosshairs denote +/- two standard errors.

Such analyses can often
be useful for assessing whether there is conflict between the marginal prior and marginal likelihood.  We note that in applications where approximation of the likelihood itself is the goal, the prior can be chosen to be whatever is convenient (in the case of approximation of the marginal likelihood, it is the marginal prior for the parameter of interest that can be so chosen).  If there is prior-likelihood conflict
then the resulting estimated likelihood may be poor, since the quality of the approximation will be very dependent on how well the tails of the posterior are estimated.  
It is an interesting question how best to choose the 
prior when the goal is likelihood approximation and we do not pursue this further here.

\subsection{Robust Bayesian variable selection}
\label{section:robust}

We consider the problem of Bayesian variable selection in regression, where the parameter of interest is a vector of binary variables indicating which covariates
are to be included in the model for the mean response.  This is a challenging problem because the parameter of interest is discrete:  
all existing ABC regression adjustment techniques are concerned with continuous
parameters \shortcite{beaumont+zb02,blum+f10,blum+nps13}.  
Further, the  marginal adjustment strategy \shortcite{nott+fms14} is difficult to apply as it needs to be implemented for each covariate model under consideration as the marginal distribution for each parameter will change conditionally on the covariates. This will rapidly become impractical as the number of covariates increases.
As a result these methods, which were responsible
 for mitigating the ABC curse of dimensionality and 
obtaining performance competitive with the ABC copula method in the multivariate $g$-and-$k$ model analysis (Section \ref{section:gandk}), are not available in this setting. 
While there is a growing literature on ABC model choice (e.g. \shortciteNP{marin+pr15}) where multinomial regression has been used to adjust model probabilities, 
such analyses have been confined to the situation where the number 
of different models is relatively small. These methods do not extend in an obvious way to problems like the one we consider here where the number of models considered is large.

We consider the US crime dataset of \citeN{ehrlich73} in which the response is crime rate, measured as the number of offenses per 100,000 population, for 47 different US states in 1960, and  there are 15 covariates,
some of which are highly collinear.  Choice of which covariates to include gives a model selection problem with $2^{15}$ distinct models.  
Suppose that $y=(y_1,\ldots,y_n)^\top$ is the vector of responses and $X$ is the $n\times16$ design matrix (with the first column containing ones and 
the remaining columns containing the centred and standardized covariates).  Write $\gamma=(\gamma_1,\dots,\gamma_{15})^\top$ as a vector of binary indicators, where $\gamma_i=1$
means that covariate $i$ is included in the model and $\gamma_i=0$ otherwise (the intercept is always included), and define $X_\gamma$ to be the corresponding design matrix containing only those covariates included in the model as indicated by $\gamma$.

We consider the linear model
$$y=X_\gamma \beta_\gamma+\epsilon,$$
where $\beta_\gamma$ is the vector of regression coefficients in model $\gamma$ (similarly considered as a sub-vector of the full model coefficients $\beta=(\beta_0,\beta_1,\ldots,\beta_{15})^\top$) and $\epsilon\sim N_n(0,\sigma^2 I_n)$ is a vector of independent zero mean normal residuals 
with variance $\sigma^2$.  We follow a common prior specification for this framework (e.g. \shortciteNP{kohn+sc01}) and set  a beta-binomial prior on the number of active covariates in the model i.e. $P(\gamma_i=1|p_\gamma)=p_\gamma$ independently for each $i$, and $p_\gamma\sim \mbox{Beta}(a,b)$ with $a=2$, $b=10$.  
We adopt the $g$-prior of \citeN{zellner86} so that $\beta_\gamma|\gamma,\sigma^2\sim N(0,n\sigma^2 (X_\gamma^\top X_\gamma)^{-1})$
and assume that $\sigma^2\sim\mbox{InverseGamma}(a_\sigma,b_\sigma)$, with $a_\sigma=5$, $b_\sigma=5\times 200^2$ which is a fairly diffuse prior
centred on a reasonable prior guess for the residual standard deviation.  With these priors $(\beta_\gamma,\sigma^2)$ can be integrated out of the model (e.g. \shortciteNP{kohn+sc01}) to give the marginal posterior
$ \pi(\gamma|y) \propto L(y|\gamma)p(\gamma)$
with
\[
 L(y|\gamma)  \propto (n+1)^{-q_\gamma/2} \left(2b_\sigma+y^\top y-\frac{n}{n+1}y^\top X_\gamma (X_{\gamma}^\top X_\gamma)^{-1} X_\gamma^\top y\right)^{-\left(a_\sigma+\frac{n}{2}\right)},
 \]
where $q_\gamma$ denotes the number of columns of $X_\gamma$.  For the US crime dataset, the number of predictors is small enough to permit enumeration of all posterior probability of all models ($2^{15}=32,768$).  
The ten highest posterior probability models for this data set are listed in Table \ref{postprobs} (column 1).

\begin{table}[tbh]
\centering
\vspace{5mm}
\begin{tabular}{lcccc}
Exact probability      & Standard ABC       & Copula ABC & Exact  & Copula ABC  \\ 
(no outlier) & (no outlier) & (no outlier)  & (with outlier) & (with outlier) \\
\hline
$x_3, x_4, x_{13}$                                    &--&  $\checkmark$  &       --                  & $\checkmark$  \\
$x_1, x_3, x_4, x_{13}$                            &--&  $\checkmark$  &         --               &  $\checkmark$ \\
$x_3, x_4, x_{13}, x_{14}$                       &--&  $\checkmark$ &           --              &  $\checkmark$ \\
$x_1, x_3, x_4, x_{13}, x_{14}$              & --&  $\checkmark$ &           --              &  $\checkmark$ \\
$x_4, x_7, x_{13}$                                    & --&    --                       & $\checkmark$ & -- \\
$x_1, x_3, x_4, x_{11}, x_{13}, x_{14}$ &--& $\checkmark$ &             --             &  $\checkmark$ \\
$x_4, x_{13}$                                             &--&           --               &             --             & -- \\
$x_1, x_3, x_4, x_{11}, x_{13}$              & --& $\checkmark$ &              --           &  $\checkmark$ \\
$x_4, x_7, x_{13}, x_{14}$                        &--&     --                     &            --             &  --\\
$x_3, x_5, x_{13}$                                     &--& --&-- & -- \\
\hline
\end{tabular}
\caption{Ten highest posterior probability models for the US crime dataset (column 1). Checkmarks ($\checkmark$) indicate those models also selected in the top 10 based on standard ABC and a copula ABC posterior approximation. Analyses are repeated with the dataset modified to include an influential outlier.
}\label{postprobs} 
\end{table}

So far ABC methods have played no role in this analysis since the marginal likelihood for $\gamma$ is directly computable.  
However, we may use ABC to compute an approximation to the posterior distribution, $\pi(\gamma|s)\propto L(s|\gamma)p(\gamma)$,
which is conditional on a summary statistic $s$, constructed so that it's distribution is insensitive to violations of the model assumptions
in the full data model $\pi(\gamma|y)$. 
In particular, we select the summary statistics $s$ to produce robust point estimates of $\beta_\gamma$, which leads to a Bayesian variable selection framework which is insensitive to outliers.
In general the sampling distribution of the robust summary statistic is intractable, even though the likelihood for the full data $y$ is tractable, and so ABC methods are needed.
For more detailed discussion of the benefits of using insufficient statistics in order to robustify Bayesian analyses
see e.g. \shortciteN{lewis+ml14}.

In order to estimate $L(\gamma|s)$ via the copula approach, we  first estimate $\pi(\gamma_i|s)$ and $\pi(\gamma_i,\gamma_j|s)$ using ABC for each parameter $i$ and parameter pair $(i,j)$. 
Parameter specific summary statistics  are constructed as
$s_{(i)}=T_{1i}$,  the robust partial $t$-statistic for significance of covariate $i$ in the full model 
(computed using the robust regression method implemented in the {\tt lmrob} function in the R package {\tt robustbase} \shortcite{rosseuw+ctrsvkm15} with the \citeN{koller+s11} method).
In addition, we fit a reduced model including the covariates 
$x_1, x_3, x_4, x_{11}, x_{13}$ and $x_{14}$ --  a ``good" reduced model for the observed data which contains only one covariate from any pair of covariates that are
highly correlated.  Then for $i\in G=\{1,3,4,11,13,14\}$ 
the robust partial $t$-statistic for the corresponding variable in this reduced model, $T_{2i}$, is added to the summary statistic vector that is informative for $\gamma_i$. That is, $s_{(i)}=(T_{1i},T_{2i})$ for $i\in G$ and $s_{(i)}=T_{1i}$ otherwise. As before, we construct $s_{(i,j)}$, the vector of statistics informative for $(\gamma_i,\gamma_j$), as the union of the marginally informative vectors $s_{(i)}$ and $s_{(j)}$.

The final estimates of $\pi(\gamma_i|s)$ and $\pi(\gamma_i,\gamma_j|s)$ are determined via ABC, by generating $N=100,000$ samples $(\gamma^{(\ell)},s^{(\ell)})\propto L(s|\gamma)p(\gamma)$ from the prior predictive distribution, and retaining the $n=n'=500$ samples closest to the observed summary statistics using Euclidean distance and a uniform smoothing kernel $K_h(\cdot)$. 
The frequency of $\gamma_i=1$ within these $500$ samples provides an estimate of $P(\gamma_i=1|s)$ and hence  an
estimate $\hat{\pi}(\gamma_j|s_{(i)})$
of $\pi(\gamma_i|s)$. Similar estimates $\hat{\pi}(\gamma_i,\gamma_j|s_{(i,j)})$ can be obtained for the bivariate posterior distribution $\pi(\gamma_i,\gamma_j|s)$.

In the discrete setting, a Gaussian copula model for $\gamma$ is defined via a latent Gaussian variable $Z=(Z_1,\dots,Z_{15})^\top \sim N(0,\Lambda)$ where $\Lambda$ is a correlation matrix.  By setting $\gamma_i'=I(Z_i>\Phi^{-1}(p_i))$, where $I(\cdot)$ is the indicator function and 
$p_i=\hat{\pi}(\gamma_i=0|s_{(i)})$, for $i=1,\dots,15$, 
then the marginal
distribution of $\gamma_i'$ is that of $\hat{\pi}(\gamma_i|s_{(i)})$. The correlation matrix $\Lambda$ can similarly be chosen so that the joint distribution of $(\gamma_i',\gamma_j')$ is
that of $\hat{\pi}(\gamma_i,\gamma_j|s_{(i,j)})$.  In particular, $\Lambda_{ij}$ is chosen so that
$$\int_{\Phi^{-1}(p_j)}^\infty \int_{\Phi^{-1}(p_i)}^\infty \phi(z_i,z_j;\Lambda_{ij})dz_i \,dz_j = \hat{\pi}(\gamma_i=1,\gamma_j=1|s_{(i,j)}),$$
where the solution for $\Lambda_{ij}$ of this nonlinear equation can be obtained numerically.  
Once the copula parameters have been estimated, joint posterior model probabilities $\pi(\gamma|s)$ for any desired value of $\gamma$ can be estimated via the copula approximation.

The middle column of Table \ref{postprobs} indicates which of the ten highest (exact) posterior model probability models conditional on the full data $y$, are also among the ten highest posterior model probabilities under the ABC copula approximation.
Six out of the exact top ten models are correctly identified as being in the top ten using the copula ABC approach. In contrast, when performing standard ABC using the full 21-dimensional vector of summary statistics (i.e. constructed as the union of the statistics in $s_{(1)},\ldots,s_{(15)}$), none of the exact top ten models are identified. In fact, the top ten models under standard ABC consist of the null model, and 9 models with a single predictor. As these top posterior models effectively coincide with the top models {\em a priori}, as the beta-binomial prior essentially favours models with fewer predictors, this indicates that the standard ABC posterior approximation is very poor, particularly in comparison with the copula ABC approximation.

As the original motivation for using ABC was to obtain a method for robust regression,
we now modify one of the observations so that it is an extreme outlier.  In particular, for the original dataset we modify the last
response value by increasing its residual standard error estimate (based on the {\tt lmrob} fit for the full model) by a factor of $10$.  
The last two columns in Table \ref{postprobs} indicate which of the original exact ten highest posterior probability models are still among the ten highest posterior probability models
when using the modified dataset.  For the exact model probabilities, conditioning on $y$, the non-robustness of the regression model to outliers is apparent as only one of the original models are still among the ten highest exact posterior probability models.  
However, for the robust ABC estimates of the model probabilities, both with and without outliers, the same 6 models remain in common with the ten best models in the exact analysis 
without outliers.  Clearly, the copula ABC method conditioning on robust summary statistics seems useful for 
finding a set of good high posterior probability models in datasets which might be contaminated by a small number of outliers.  
Equally clearly, standard ABC methods are not useful for this purpose.

\section{Discussion}
\label{sec:discussion}

The standard construction of ABC methods, based on conditional kernel density estimation, means that they do not extend well to high dimensional analyses due to a curse of dimensionality on the vector of summary statistics, $s$.
The copula approach introduced in this paper constructs a Gaussian copula approximation to the full ABC posterior distribution. In this manner, the need to simultaneously match a high-dimensional vector of simulated and observed summary statistics is circumvented in favour of separately matching many low-dimensional vectors to form the copula approximation. The fitted copula is not always appropriate to approximate certain highly complex posterior distributions, as it assumes a Gaussian dependence structure (i.e. based on bivariate linear correlations), albeit with flexible univariate marginal distributions. 
This means that non-linear dependencies, or complex higher-order relationships between three or more parameters in the full posterior $\pi(\theta|s_{obs})$ will not be accurately captured.
However, copula ABC may be adequate in many modelling situations, 
especially those where an accurately fitted Gaussian copula approximation to a highly complex posterior may be more practically useful than a very poor standard ABC approximation to the joint model (see e.g. Sections \ref{section:toy} and \ref{section:robust}).
The copula structure will also become a more appropriate approximation to the true posterior as the sample size increases, and the true posterior approaches normality.
As such, copula ABC is a useful and viable general technique for directly extending ABC modelling to  high-dimensional problems.

One point of practical consideration for copula ABC is the requirement to select $s_{(i)}$ and $s_{(i,j)}$ i.e. those subsets of $s$ that are informative for $\theta_i$ and $(\theta_i,\theta_j)$. 
In principle, this could take the same amount of work
in identifying the vector $s$ that is informative for $\theta$, but repeated many times, over each univariate and bivariate posterior margin. The semi-automatic work of \citeN{fearnhead+p12} is useful here, in that it provides a principled way of identifying linear combinations of the elements of a vector of summary statistics that are informative for a subset of parameters (they are in fact, Bayes linear estimates of those parameters; \shortciteNP{nott+fs12}). While it should be noted that these semi-automatic statistics are only optimal for the posterior mean, rather than any measure on the joint distribution, they have been successfully implemented in a large range of applications. Beyond this, the analyst can alternatively make use of knowledge of the structure of the model in order to identify informative subsets of $s$. We used this approach (in combination with the semi-automatic approach) with each of our analyses in Section \ref{sec:examples}, although an alternative would have been to use the semi-automatic approach directly for each bivariate margin. In general, the principled identification of summary statistics for ABC methods remains a challenging practical problem (e.g. see \shortciteNP{blum+nps13}).

As well as improving estimation of the posterior dependence structure,
copula ABC may also be very valuable because it provides an approximate analytic expression for the posterior density.  As previously discussed, 
this can be used to build a likelihood approximation, and permit frequentist analyses that can serve as a reference for comparison with a Bayesian analysis.  Approximation of 
likelihood functions can also be important in the context of setting informative priors in fully Bayesian analyses, for example in the so-called power prior approach (\shortciteNP{chen+i00}). 
Here, a tempered version of the likelihood for past, indirect data $z$ is used to set the prior for the analysis of the current data, $y$.  Even when the likelihood for the data $y$ is tractable, our knowledge of the past data $z$ might be limited to summaries for which the corresponding likelihood is not tractable.  Our copula ABC approach would then provide a way to make the required likelihood approximations for the past data in this situation.

\section*{Acknowledgements}

DJN is supported by a Singapore Ministry of Education Academic Research Fund Tier 2 grant (R-155-000-143-112). SAS is supported by the Australian Research Council through the Discovery Project scheme (DP160102544).

\bibliographystyle{chicago}
\bibliography{abc-biblio}

\end{document}